\documentstyle[11pt,newpasp,twoside,psfig]{article}
\def\edcomment#1{\iffalse\marginpar{\raggedright\sl#1\/}\else\relax\fi}
\reversemarginpar

\def\kms{km~s$^{-1}$}

\begin{document}
\title{Anomalous X-ray pulsars and soft gamma-ray repeaters in 
supernova remnants}
\author{B. M. Gaensler\altaffilmark{1}}
\affil{Center for Space Research, Massachusetts Institute of Technology, \\
70 Vassar Street, Cambridge, MA 02139, USA; bmg@space.mit.edu}
\altaffiltext{1}{Hubble Fellow}

\begin{abstract}
I consider the state of play regarding associations of
supernova remnants (SNRs) with anomalous X-ray pulsars (AXPs)
and soft gamma-ray repeaters (SGRs). The three AXP/SNR associations
are convincing, and are consistent with AXPs being young, low-velocity
neutron stars. The three SGR/SNR associations are far more likely
to be chance superpositions, and rely on SGRs being high velocity
($>$1000~\kms) objects. These results imply either that AXPs evolve
into SGRs, or that SGRs and AXPs represent different populations
of object.
\end{abstract}

\section{Introduction}

The recent detection of rapidly slowing $\sim$6-second pulsations
from soft gamma-ray repeaters (SGRs) makes a strong argument that
these sources are {\em ``magnetars''}, isolated neutron stars
with inferred dipole magnetic fields $B \sim 10^{14} - 10^{15}$~G
(e.g.\ Kouveliotou et al. 1998).

Thompson \& Duncan (1996) have noted that the emergent class of
six {\em ``anomalous X-ray pulsars''} (AXPs; van Paradijs et al.\ 1995)
are strikingly similar to SGRs in their periods, period
derivatives, X-ray luminosities, X-ray spectra,
lack of evidence for binarity and coincidence with supernova
remnants (SNRs). They 
thus propose that AXPs, like SGRs, are magnetars. In the
subsequent few years, several more AXPs and SGRs have been discovered,
several of which are near or in SNRs (e.g.\ Vasisht \& Gotthelf 1997;
Woods et al. 1999; Gaensler et al. 1999). Below I briefly summarise these
associations, then consider what these results tell us about AXPs, SGRs,
and the relationship between the two populations.

\section{Associations of SNRs with AXPs and SGRs}

Claimed associations of  SNRs with AXPs and SGRs are summarised in Table~1.
Note that the association between SGR~1806--20 and
G10.0--0.3 (Kulkarni et al. 1994) has been omitted, 
as the latter appears to be a synchrotron nebula
powered by the SGR (or perhaps by some other source; Eikenberry,
these proceedings) and gives no evidence for a supernova explosion
at some point in the past.

For each association, I have listed an estimated age  and distance
for the SNR. It should be noted that the $\Sigma-D$ relation
is not a valid method of determining distances to individual
SNRs (e.g.\ Green 1984),
and that distances derived using this method
should not be taken seriously. Age estimates
for SNRs are also uncertain, and usually depend on assumptions about
the ambient density.

The parameter $\beta$ corresponds to the offset of a compact object
from the apparent centre of its SNR, in units of the SNR radius (e.g.\
Shull et al. 1989). For example,
$\beta = 1$ corresponds to an AXP or SGR sitting on the rim
of its associated SNR. The column $V_T$ refers
to the implied transverse velocity of the pulsar, using the adopted age,
distance and offset.

\begin{table}
\caption{Associations of SNRs with AXPs and SGRs. Values in {\em italics}\
are representative, and do not correspond to measured quantities.}
\label{tab_snrs}
\begin{tabular}{llccccl} \tableline
AXP/SGR   & SNR  &  $t_{\rm SNR}$ & $d_{\rm SNR}$ & $\beta$ & $V_T$ & Ref. \\
         &      &  (kyr)          & (kpc) &        & (\kms) & \\ \tableline
1E~1841--045 & Kes~73 & 2 & 7 & $<$0.2 & $<$500  & 1, 2 \\
AX~J1845--0258 & G29.6+0.1 & $<$8 & {\em 20} & $<$0.15 & $<$500 & 3 \\
1E~2259+586 & CTB~109 & $\sim$10 & 5 & $<$0.25 & $<$500 & 4, 5  \\ \tableline
SGR~0526--66 & N~49 & 5 & 50 & 1 & 2900 &   6 \\
SGR~1627--41 & G337.0--0.1 &  5 & 11  & 2 & 800 & 7 \\
SGR~1900+14 & G42.8+0.6 & {\em 10} & {\em 10} & $\sim$1.2 & $\sim$1800 & 
--- \\
\tableline \tableline
\end{tabular}

References for ages \& distances: 
(1) Sanbonmatsu \& Helfand (1992); (2) Vasisht \& Gotthelf (1997); (3) 
Gaensler et al. (1999); (4) Green (1989); (5) Rho \& Petre (1997);
(6) Vancura et al. (1992); (7) Corbel et al. (1999)
\end{table}

\subsection{Anomalous X-ray Pulsars}

Associations between neutron stars and SNRs are usually judged
on criteria such as agreement in age/distance, positional
coincidence and evidence from proper motion. Distance estimates
for AXPs have uncertainties $\ga$50\%, and there is no evidence
that
their characteristic ages ($\tau_c \equiv P/2\dot{P}$)
are reliable age estimators. We also lack proper
motion measurements for these sources, and so are left only with
positional coincidence in order to judge associations.

In all three cases in Table~1, the AXP is sitting almost exactly
at the centre of its SNR. The probability of random superposition
is thus very small, $<$0.2\% (see Gaensler et al. 1999), and we can conclude
that all three AXPs are likely to be physically associated with
their coincident SNRs. The upper limits on the AXPs' transverse
velocities are entirely consistent with the
velocity distribution seen for radio pulsars (e.g. Lyne \& Lorimer 1994). 
Both the
ages of the associated SNRs, and the values of $\beta$ argue
strongly that AXPs are young ($<$10~kyr) objects;
the apparent absence of SNRs around the
remaining three AXPs is consistent with the expectation
that many (or even most) SNRs occur in low density regions,
and do not produce detectable emission (Kafatos et al. 1980; Gaensler \&
Johnston 1995b). This result 
implies a Galactic 
birth-rate for AXPs of $>$0.6 kyr$^{-1}$, corresponding to at least
5\% of core-collapse supernovae (see Gaensler et al. 1999).

\subsection{Soft Gamma-ray Repeaters}

Just as for the AXPs, we cannot appeal to age, distance
or proper motion in considering associations between SGRs and SNRs.
Turning to positional coincidence, we find that all three SGRs are
on the edge of, or outside, their coincident SNRs. The probability
of a chance coincidence increases as $\beta^2$, and one consequently
finds a substantially higher probability than for the AXPs that the
SGR/SNR associations are spurious (e.g. Smith et al. 1999).
Of the $\sim$10 claimed associations between SNRs and 
radio pulsars with $\beta>1$, 
all but one is likely to be geometric projection
(e.g. Gaensler \& Johnston 1995a,b;
Nicastro et al. 1996; Stappers et al. 1999).

Thus we are left to conclude either that the SGR/SNR associations
are not genuine, or that SGRs have substantially higher velocities
than do radio pulsars. 
There is currently no way to distinguish
between these possibilities; using {\em Chandra}\ to measure
the proper motion of the SGRs seems to be the only avenue
by which this might be resolved.
We note that
Duncan \& Thompson (1992) argue that the mechanism which
forms a magnetar will indeed impart the neutron star with a
high recoil velocity, consistent with the values of $V_T$
for the SGRs in Table~1. 

\section{Relationship between AXPs and SGRs}


On the basis of the small value of $\tau_c$ for
SGR~1806--20, Kouveliotou et al. (1998) have
argued that SGRs eventually evolve into AXPs.
Meanwhile, Gotthelf et al. (1999) appeal to the young age
of the Kes~73/1E~1841--045 association to argue
that AXPs evolve into SGRs!
However, if all the associations in Table~1 are genuine,
then AXPs and SGRs clearly have different velocity distributions and
so cannot possibly be drawn from the same population, coeval or otherwise.


On the other hand,
if one argues that the SGR/SNR associations in Table~1 are
merely chance coincidence, then the corresponding estimates
of $V_T$ are invalidated.  The
absence of associated SNRs for SGRs would then imply that SGRs have
ages $\ga$50--100~kyr (e.g.\ Shull et al. 1989; Frail et al. 1994), 
and the data would then be consistent with AXPs evolving into SGRs.
One possible problem with this scenario is that 
if one extrapolates the steady spin-down seen in several
AXPs to such ages (Gotthelf et al. 1999; Kaspi et al. 1999),
we would then expect SGRs to have periods $\gg$10~s, 
which is not observed. 

\section{Conclusions}

The three associations between AXPs and SNRs are all convincing,
and indicate that AXPs are young ($<$10~kyr), low velocity
neutron stars. The three SGR/SNR associations seem less likely
to be genuine, and rely on SGRs being high
velocity ($>$1000~\kms) objects. 
If the SGR/SNR associations are indeed spurious, then SGRs
can be explained as older manifestations of AXPs.
However, if the SGR/SNR
associations are shown to be real, then we must
conclude that there is no evolutionary link between
SGRs and AXPs. Possible alternatives are that
AXPs are accreting systems as originally claimed (e.g.
van Paradijs et al. 1995), or that there is more than one type of magnetar.

\acknowledgments
My research is supported by
NASA through Hubble Fellowship grant
HF-01107.01-98A awarded by the Space Telescope Science Institute, which
is operated by the Association of Universities for Research in
Astronomy, Inc., for NASA under contract NAS 5-26555.


\begin{references}
\reference Corbel, S., Chapuis, C., Dame, T. M.,
\& Durouchoux, P. 1999, \apjl, 526, L29
\reference Duncan, R. C., \& Thompson, C. 1992, \apjl, 392, L9
\reference Frail, D. A., Goss, W. M., \& Whiteoak, J. B. Z. 1994,
\apj, 437, 781
\reference Gaensler, B. M., Gotthelf, E. V., \& Vasisht, G. 1999,
\apjl, 526, L37
\reference Gaensler, B. M., \& Johnston, S. 1995a, PubASA,
12, 76
\reference Gaensler, B. M., \& Johnston, S. 1995b, \mnras, 277, 1243
\reference Gotthelf, E. V., Vasisht, G., \& Dotani, T. 1999, \apjl,
522, L49
\reference Green, D. A. 1984, \mnras, 209, 449
\reference Green, D. A. 1989, \mnras, 238, 737
\reference Kafatos, M., Sofia, S., Bruhweiler, F., \& Gull, T.
1980, \apj, 242, 294
\reference Kaspi, V. M., Chakrabarty, D., \& Steinberger, J. 1999,
\apjl, 525, L33
\reference Kouveliotou, C. et al. 1998, Nature, 393, 235
\reference Kulkarni, S. R. et al. 1994, Nature, 368, 129
\reference Lyne, A. G., \& Lorimer, D. R. 1994, Nature, 369, 127
\reference Nicastro, L., Johnston, S., \& Koribalski, B. 1996,
\aap, 306, L49 
\reference Rho, J., \& Petre, R. 1997, \apj, 484, 828
\reference Sanbonmatsu, K. Y., \& Helfand, D. J. 1992, \aj,
104, 2189
\reference Shull, J. M., Fesen, R. A., \& Saken, J. M. 1989,
\apj, 346, 860 
\reference Smith, D. A., Bradt, H. V., \& Levine, A. M. 1999,
\apjl, 519, L147
\reference Stappers, B. W., Gaensler, B. M., \& Johnston, S. 1999,
\mnras, 308, 609
\reference Thompson, D., \& Duncan, R. C. 1996, \apj, 473, 322
\reference van Paradijs, J., Taam, R. E., \& van den
Heuvel, E. P. J. 1995, \aap, 299, L41
\reference Vancura, O., Blair, W. P., Long, K. S., \& Raymond, J. C.
1992, \apj, 394, 158
\reference Vasisht, G., \& Gotthelf, E. V. 1997, \apjl, 486, L129
\reference Woods, P. M. et al. 1999, \apjl, 519, L139
\end{references}
\end{document}